\documentstyle[11pt]{article}
\title{Condensation and Metastability in the 2D Potts Model}
\author{\bf J. L. Meunier$^{1,a}$ and A. Morel$^{2,b}$}
\begin{document}
\input{psfig}
\maketitle

\centerline{$^1$Institut non Lin\'eaire de Nice-Sophia Antipolis, 1361 Route des Lucioles,}
\centerline{06560 Valbonne, France} 
\vspace{0.3cm}

\centerline{$^2$Service de Physique Th\'eorique de Saclay, CE-Saclay,} 
\centerline{F-91191 Gif-sur-Yvette Cedex, France} 
\vspace{-12cm}

\begin{flushright}
SPhT-99/030 \\
INLN-99/003

\end{flushright}

\vspace{10cm}

\begin{abstract} \normalsize
{For the first order transition of the Ising model below $T_c$, 
Isakov has proven that the free energy possesses an essential 
singularity in the applied field. Such a singularity in the control 
parameter, anticipated by condensation theory, is believed to 
be a generic feature of first order transitions, but too weak to be
observable. We study these issues for the temperature driven 
transition of the $q$ states $2D$ Potts model at $q>q_c=4$. Adapting the
droplet model to this case, we relate its parameters to the
critical properties at $q_c$ and confront the free energy to
the many informations brought by previous works.
The essential singularity predicted
at the transition temperature leads to observable effects
in numerical data. On a finite lattice, a metastability domain of
temperatures is identified, which shrinks to zero 
in the thermodynamical limit.  }
\end{abstract}

\vspace{0.5cm}

\noindent
$^a$meunier@inln.cnrs.fr \\
$^b$morel@spht.saclay.cea.fr
\section{Introduction}
The transition between the vapour and liquid phases of a 
material has been for a long time the prototype for first order transitions. 
Below a critical temperature $T_c$, the density, that is the derivative of the 
pressure $\cal P$ with respect to the chemical potential $\mu$, 
is discontinuous at $\mu =\mu _t$, where the two phases coexist. The 
vapour phase is stable below $\mu _t$, the liquid phase above.
As $T \to T_c$, the transition becomes continuous, the density fluctuations 
diverge, leading to the well known phenomenon of critical opalescence. 
Van der Waals theory of condensation predicts these equilibrium properties, 
and in addition lead to the existence of metastable states, 
vapour above $\mu _t$ and liquid below, experimentally accessible
during finite times. In this theory, they correspond to the possibility
of analytically continuing the low  and
high density branches of the isotherm function ${\cal P}(\mu )\left|_T\right.$
beyond $\mu _t$.

All of this can be found in standard textbooks,
together with other methods leading to similar descriptions, as well
as applications of the same ideas to many systems of physical and/or 
theoretical interest, which undergo first order transitions. One of the
simplest and most extensively studied system is the Ising model.
There instead of the pressure, one considers ${\cal F}(h)\left|_T\right.$, the free energy as a function of the applied magnetic field $h$ at fixed
$T$. Below $T_c$, the field driven first order transition is located
at $h_t=0$, and manifested by a discontinuity of the magnetization. As
$T \to T_c$, the magnetization vanishes, and its fluctuations measured
by the susceptibility, diverge with a critical exponent which is 
known in 2D. In this particular case however, it has been
mathematically proven by Isakov \cite{isakov} that Van der Waals
theory cannot be true: each of the ordered phase free energies possesses an {\it essential
singularity at $h=h_t$}, which makes its analytical continuation 
ambiguous in the metastability domains where the magnetization remains 
opposite to the applied field.

In fact, such a singularity in the control parameter at 
the transition was anticipated a long time ago from the
droplet theory of condensation in the version developped by Fisher 
\cite{fisher}, who attributes to Mayer \cite{mayer} the first
suggestion in this direction. Shortly after, Langer
\cite{langer} applied the droplet model to the Ising case, emphasizing
that the expected singularity was so weak that it could not be experimentally
detected. The reader will find in \cite { fisher,langer} many 
references to earlier and related work, and inspiring discussions on the 
physics behind. Since then, the case has been reexamined many times,
with particular concerns about the definition and properties of the
so-called droplets, or clusters. Steps in this line of research
can be followed in, e.g., \cite {binder, delyon, aizenmann}. Its
intimate connection with metastability and dynamics at first order
transitions has been for a long time a matter of investigations and
debate \cite {gunton, binderrev}, and 
still is an active field \cite{schulman, dynamics, abraham, rikvold, gunther, 
acharyya}.

The basic idea of the
approach is that the fluctuations of the system in a stable phase
are due to the appearance of droplets of one or several other phases, 
whose size and number are controlled by the value of the driving
parameter. The statistics of these droplets is described as a 
Mayer's cluster expansion of the thermodynamical function at hand, 
and the analyticity of the latter follows from the 
convergence properties of this expansion. A transition occurs at a point
of parameter space where the energy/entropy balance favours 
large clusters of the wrong phase, and the cluster expansion is shown
to diverge at such a point, which thus is singular. 

In this paper, we investigate similar issues within the bidimensional, 
$q$-states, Potts model \cite{potts} for which we adapt the droplet
picture, much in the spirit of \cite{fisher}, using 
the fact that the transition, here driven by temperature
and being first order for $q>4$, becomes continuous as $q\to q_c=4$
where the latent heat
and the interfacial tension vanish.
The Potts model is interesting per se, as illustrated by the many
studies which have been and still are devoted to it, and it seems thus
worthwhile to learn more about it. As we shall see, the droplet
point of view allows one to put together in a very compact way many of the
results accumulated over years, either analytically or by high precision 
 simulations \cite{jknum, billoire} at the transition. It
is the case for the relationship between properties of the free energy along
the first order transition line $q>4$ and at its critical end point $q_c$,
a relationship which was evidenced in a series of papers by Bhattacharya
et al. \cite {bhat1,bhat2,bhat3}. This connection was recently studied in a
completely different framework by Cardy \cite{cardy} who relates the
emergence of a critical $q_c$ to branching properties of the Potts
interfaces between different ordered phases.  A better knowledge of 
the analytic structure of the free energy close to a first order point may also help in other domains of current interest, such as the dynamics, as
already mentioned, and the role of disorder
for the phase diagram of statistical systems.

Duality relates the disordered phase, stable at high
temperature, to the $q$ ordered phases, stable at low temperature. 
We will mainly focuss on the disordered phase. The paper is organized as follows. In Section (2), we adapt the
droplet formulation to the Potts case, and constrain its
parameters by requiring consistency, for $q$
values  above 4, with the known critical properties {\it at} $q_c$.
In Section (3), we first show that, in a pure phase, the order $n$ energy cumulants $f_n$
of the model compare successfully with those measured in \cite{jknum} for
$n\leq 8$ and $q= 10, 15, 20$. Only one, $q$-independent, parameter is
needed to reproduce all ratios $f_n/f_2$ given $f_2$ at each $q$. 
This scaling property allows one to construct a universal
(i.e. $q$-independent) function $\phi (z)$, which represents the pure
phase free energy suitably rescaled, $z$ measuring the distance 
to the transition point in units of the inverse temperature $\beta$, also
rescaled. The function $\phi$ has a simple integral representation from
which we derive its analytic properties. It has an essential 
branch point singularity at $z=0$. We then turn to applications. The 
associated internal energy distribution for a finite square lattice is 
studied in Section (4). In particular,
rescaling the internal energy and the lattice size leads to a
universal distribution from which any $q$-dependence has disappeared.
We emphasize consequences for the distribution 
of the singular structure of the free energy at the transition.
Specific finite size effects are predicted, {\it although by assumption 
the thermodynamical limit of the free energy density has been taken}, 
and compared with those actually 
observed in a numerical simulation performed at $q$=9. In Section
(5), we define the disordered distribution {\it below} the transition 
temperature by reweighting that defined above. From its finite size 
properties, we determine a size dependent spinodal value $\beta ^*$ such
that the system is metastable in the interval $\beta ^*>\beta >\beta _t$.
Inside this interval, which shrinks to zero in the thermodynamical
limit, we define a free energy for metastable states .
A summary and conclusions are proposed in a last section.

\section{The Potts model and its droplet description}
We first recall a few basic definitions and properties relevant
for our discussion of thermodynamical quantities associated with
the ferromagnetic $q$-states Potts model. 
We next summarize the standard droplet description of condensation, and finally 
adapt it so as to incorporate the known properties of the model at 
$q\ge 4$ .
\subsection{The Potts Model}
The model is defined \cite{potts}through the Hamiltonian
\begin{equation} \label{H}
H=-J\sum_{<ij>}\delta_{{\sigma_i}{\sigma_j}},\
\end{equation}
where $<ij>$ denotes the pairs of nearest neighbours on a square lattice
of area $A=L^2$ and $\sigma_i$ one of the $q$ possible values of the 
spin variable at site $i$. In
the rest of the paper, the energy per link $J$ is set equal to 1.\\
The partition function of the system is 
\begin{eqnarray} 
Z_A(\beta )&=&\sum_{\{\sigma\}}\exp(-\beta H({\sigma})),\\ 
&=&\sum _E\Omega _{A}(E)\exp{(-\beta A\,E)} \label{ZA}
\end{eqnarray}
where $\beta$ is the inverse temperature, $E$ $\in [-2,0]$ the  
energy density per site, and $\Omega _A$ the corresponding number of states. 
It is equivalent to say that the energy probability density is
\begin{equation} \label {pb}
P_{\beta ,A}(E)=\frac {1}{Z_A(\beta )}\,\,\Omega _{A}(E)\,\exp (-\beta A\,E).
\end{equation}
For convenience, we shall call "free energy" the quantity
\begin{equation} \label{FA}
F_A={1\over A}\log (Z_A), \ 
\end{equation}
while strictly speaking $F_A$ is $-1/\beta$ times the standard free 
energy density. Its thermodynamical limit 
$F$ is its limit as $A \to \infty $. \
The finite size microcanonical entropy density $S_A(E)$ and its
thermodynamical limit $S(E)$ are defined by
\begin{eqnarray} \label {entropy}
S_A(E)&\equiv &{1\over {A}}\,\log (\Omega _A ), \\
&=&F_A(\beta )+\frac {1}{A}\log (P_{\beta ,A}(E))+\beta \,E,\\
S(E)&=&lim\, S_A(E) ,\quad A\rightarrow \infty.
\end{eqnarray}
A numerical simulation provides the energy distribution $\Omega _{A}
e^{-\beta A\,E}$ up to a numerical factor, and thus the entropy $S_A$
up to an additive constant.\\
A review of many of the known properties of the Potts model can be found
in \cite{wu}. A transition occurs at an inverse temperature $\beta _t$
given by \cite {baxter}
\begin{equation} \label{bet}
{\beta_t}=log(\sqrt q+1). \
\end{equation}
It is first order for $q>4$, and second order at and below 
the end point $q_c=4$, where it is characterized, e.g., by the critical indices 
$\alpha$ and $\nu$ for the specific heat and the correlation length 
respectively . For future reference, we recall that 
\begin{eqnarray} \label{critind}
\alpha &=&2/3, \nonumber \\
\nu &=&2/3. \
\end{eqnarray}
In the thermodynamical limit, the system is either in
the disordered phase, for $\beta <\beta _t$, or in one of the $q$ degenerate 
ordered phases for $\beta >\beta _t$. When necessary, a quantity
referring to an ordered or to the disordered phase will receive a superscript
$o$ or $d$ respectively. So $F^{(o)}$ (resp. $F^{(d)}$) denotes the ordered
(resp. disordered) free energy, a well defined function of
$\beta$ for $\beta > \beta _t$ (resp. $<\beta _t$). The transition point is the
$\beta$ value where the two free energies are equal: $F^{(o)}(\beta _t)=F^{(d)}(\beta _t).$ For $q>4$ where the
transition is first order, one expects on very general grounds that at
$\beta_t$,
$F^{(o)}$ (resp. $F^{(d)}$) has finite left (resp. right) derivatives to any
order $n$ with respect to $\beta$. They give ${(-1)}^n$ times the  cumulants $f_n^{o,d}$ 
of the internal energy associated with each of the
pure phases $o$ or $d$. We shall be 
interested in the analytic continuation of the free energies in the complex 
$\beta$ plane at  $q>4$ fixed, which will consist in giving a meaning to 
the formal expansion:
\begin{equation} \label{series}
F^{(o,d)}(\beta)=\sum _{n=0}^\infty {(-1)}^nf_n^{o,d}(\beta - \beta _t)^n/n!\qquad .
\end{equation}
The two functions $F$ are related to each other by duality \cite{wu}.
If 
\begin{equation} \label{duality}
(\exp(\tilde \beta)-1)\,(\exp(\beta)-1)=q,
\end{equation}
then
\begin{equation}
F^{(o)}(\tilde \beta)=F^{(o)}(\beta)-2\,\log((\exp(\beta )-1)/\sqrt q) \quad.
\end{equation}
We choose to study $F^{(d)}$ for concretness, and consider the $d \to o$ transition
as a condensation process where  droplets of aligned spins inside a 
disordered bulk tend to grow as $\beta$ approaches $\beta _t$ from below. 
The droplet model consists in describing the system at equilibrium as 
a statistical distribution of non interacting droplets.

\subsection{The Droplet Picture}

We closely follow Fisher \cite{fisher}. The transposition from vapour
condensation to the $d\to o$ Potts transition at $q> q_c$ produces
the disordered phase free energy $F^{(d)}(\beta )$ as a function of the inverse 
temperature. A droplet of size $\ell$ (its area) is a connected domain of 
$\ell$ sites with the same spin value. The model assumes that it has an 
effective perimeter scaling as $\ell^\sigma$, where the exponent
$\sigma$, smaller than 1 (no fully ramified clusters), but possibly larger than the geometrical value 1/2 \cite{hiley}, accounts for the appearance at fixed 
$\ell$ of many different shapes and topologies. This is controversial
and calls for comments to be made later. The free energy can then be written:

\begin{eqnarray} 
F^{(d)}(\beta)&=&c\sum_{\ell=1}^\infty \ell^{-\tau}x^{\ell^\sigma}y^\ell \label{droplet}\\
y&=&{\exp (\beta-\beta _t)}\label{y}\\
x&=&{\exp (-\omega)} \label{x}\
\end{eqnarray}
with
\begin{equation} \label{inequal}
{\omega >0, \qquad 1/2<\sigma <1,\qquad   \tau >0.}
\end{equation}
The parameter $c$ fixes the normalisation and $\ell^\tau$ is a
correction to the area and perimeter dependences \cite{langer, essam}. 
A priori, the parameters $\omega, \sigma, \tau$
and $c$ are functions of $q$, a question which we examine in the next
subsection.
Eqs. (\ref{droplet}--\ref{x}) are interpreted as follows.
The free energy is the sum over $\ell$ of contributions coming from 
all clusters of size $\ell$. The quantity $y$ plays the role of the
activity in gas condensation or of $\exp {(-h)}$ in the Ising problem, and
$\omega$ denote an energy density per unit of effective perimeter (effective 
interface tension between the interior and the exterior of a droplet). 

Each term in the above sum is proportional
to the probability of an ordered cluster of size $\ell$ in the system. 
Due to inequalities (\ref{inequal}), the factor $y^\ell$ implies
that for $\beta <\beta _t$ the
probability of arbitrary large clusters is exponentially small. The disordered 
phase is stable there, the series and all its derivatives with
respect to $\beta$, converge. The converse 
is true for $\beta >\beta _t$. It follows that the disordered 
free energy $F^{(d)}$ is analytic in the half-plane $Re(\beta) <\beta_t$ and has an
essential singularity at $\beta _t$. 

As a side remark, we note that 
in its Kasteleyn-Fortuin formulation 
\cite {kasteleyn}, the Potts model is a model of satisfied or
unsatisfied links. The temperature dependence of
the partition function is carried by weigths $\bar y^l$, with 
$\bar y=(\exp(\beta)-1)/\sqrt q$, $\ell$ being the number of satisfied links
forming a connected cluster. The droplet picture applied to the
link representation leads to equations similar to (\ref {droplet},\ref
{y}) with $y$ replaced by $\bar y$. Close to the transition,
$y\simeq \bar y\simeq 1$ up to order $(\beta -\beta _t)$ terms, so that
using $y$ or $\bar y$ makes no difference for the leading singularity: 
It has the same structure in $b=\log \bar {y}$ and in $\beta -\beta _t$. 

Following Fisher \cite{fisher}, we  now consider the system close 
to the critical point $q_c$, and fix the parameters $\omega$, 
$\sigma$ and $\tau$.

\subsection{Fixing the Parameters from the Critical
Point}
We go on with the disordered phase free energy, and from now on
omit the $d$ superscript. The energy cumulant of order $n$ is given by
\begin{equation}
{f_n=(-1)^n{d^n\over d\beta^n}F\vert _{y=1}}
\end{equation}
with F given by Eq.(\ref {droplet}).
At least for large enough orders, the series obtained for $f_n$ 
can be given a closed form, by replacing the discrete
sum over $\ell$ by an integral \cite{fisher,langer}.
The result is 
\begin{eqnarray} 
f_n&=&c(-1)^n\int_0^\infty{ d\ell\exp{(-\omega \ell^\sigma})\ell^{n-\tau}}\\
&=&{{c(-1)^n}\over {\sigma \omega^{(n-\tau+1)/\sigma}}}{\Gamma{((n-\tau+1)/\sigma)}.} \label{cumul}
\end{eqnarray}

Hence all the derivatives of $F$ exist at the transition, but due to 
$\sigma <1$ the convergence radius of its Taylor series (\ref {series}) 
in $(\beta -\beta_t)$ is zero, a characteristic situation for an
essential singularity at $\beta _t$. It also follows from (\ref {cumul}) that 
the scale in $\beta -\beta _t$ relevant for the behaviour of F close to 
the transition is
\begin{equation}
\beta _0=\omega ^{1/\sigma}. \label {beta0} 
\end{equation}
On another hand, from an expansion to order $10$ in
$1/\sqrt q$ of the free energy, it was empirically found in \cite {bhat1,bhat2,bhat3},
and confirmed by expansions recently pushed to order $23$ \cite
{arisue}, that at least in a ``low'' $q$ domain, in fact extending 
up to $q\approx 30$ or more, one has approximately
\begin{eqnarray} 
f_2&\propto & \xi, \label{f2}\\
f_3/f_2&\propto &\xi^{3/2},\label{f3sf2}
\end{eqnarray}
where $\xi$, the correlation length of the Potts model, is exactly 
known \cite{xi,borgs} and grows extremely rapidly as $q$ decreases towards 
$q_c$.  In \cite{bhat1,bhat2}, a tentative explanation for such behaviours
was that {\it at} $q_c$, $\xi$ and the most singular part $F_{sing.}$
of $F$ are respectively proportional to $\vert \beta -\beta
_t\vert^{-\nu}$ and $\vert \beta -\beta _t\vert ^{2-\alpha}$, so that
Eqs.(\ref {f2},\,\ref {f3sf2}) hold there, due to (\ref {critind}), and thus 
perhaps in some neighborhood of the critical point.
Together with (\ref {cumul},\,\ref {beta0}), this invites us to
postulate
\begin{eqnarray} 
\omega &=&f/{f_2}\propto 1/\xi,\label{omega}\\
\sigma&=&2/3, \label{sigma}\\
\tau&=&7/3, \label{tau}
\end{eqnarray}
with $f$ approximately constant with respect to $q$.

Note that if $\omega$ in (\ref {droplet},\ref {x}) is interpreted 
as an effective interfacial tension between the interior and the 
exterior of a droplet, its postulated behaviour (\ref {omega}) 
fits nicely that of the true $o-d$ interface tension $\Sigma$, known to
be equal to $1/{2\xi}$ \cite {borgs}, and also determined numerically 
via a study of phase coexistence at $\beta _t$
\cite {billoire}. We remark that at $q=q_c$ the same 
behaviour $\Sigma\, \xi \to cst.$ as $\beta \to \beta _c$ follows from 
the scaling relation $\mu =\nu$, $\mu$ being the Widom exponent for 
the interface tension \cite{widom}. 
Why it also holds for $\omega \,\xi$ remains however unclear since
whereas $\Sigma$ is an energy density per unit {\it length} along a straight
boundary, $\omega$ is here related to an effective boundary whose size 
scales as a {\it length to the power $2\sigma =4/3$} (\ref {sigma}).

As discussed in details by Binder \cite{binder} for the Ising case, taking
$\sigma \ne 1/2$ is quite questionable, and in fact incorrect if one 
insists that the droplets of the model are the standard geometric clusters 
which can be drawn on a lattice. This is confirmed by the exact results 
of \cite{delyon,aizenmann} about the probability of such clusters, and of
Isakov \cite{isakov} on the cumulants of the Ising model at sufficiently
low temperature. Because their derivation is essentially of geometrical nature,
we see no reason why they should not apply to the Potts case. A physical 
interpretation of $\sigma >1/2$ has been proposed by Binder. For the time 
being, we take $\sigma =2/3$ as an effective parameter describing some 
neihborhood of the critical end point, and show that anyway it  accounts 
for many observed facts. This point will be discussed again in our 
conclusions (Section (6)).

\section{The Cumulants confronted to Numerical Data. The Free Energy }
The arguments borrowed from \cite {bhat1,bhat2} to guess
$f_3/f_2$ Eq.(\ref {f3sf2}) were furthermore applied in these 
references to cumulants of any order by successive derivations of 
$F_{sing.}\propto (\beta _t-\beta)^{2-\alpha}$ and the subsequent 
replacement of $(\beta _t -\beta)$ by a constant times $\xi ^{-1/\nu}$.
As a result, the series (\ref {series}) converged, providing 
an ansatz for $F^{(d)}$ \cite {bhat3} analytic up to a value $\beta ^*>\beta
_t$, which could be interpreted as the end of a metastability region (spinode),
in clear contradiction with {\it any} droplet approach.

\subsection{The Potts model cumulants}
Let us compare the model predictions for the energy cumulants with 
the results of the high accuracy  simulation of Janke and Kappler \cite {jknum}, who give the {\it pure phase} $f_n$ up to order 8 {\it at} $\beta _t$. 
We are concerned with $n\ge 2$ only, i.e. with those cumulants
which diverge at the critical point. Anyway, $f_0$ and
$f_1$ are exactly known \cite {baxter}, and irrelevant for the study
of a single phase. Since the normalisation 
constant $c$ remains undetermined and since $f_2$ is supposed to carry most of
the $q$-dependence of $\omega$ (\ref {omega}), we rewrite (\ref {cumul}) as
\begin{equation}
{f_n\over {f_2}}=(-1)^n\bigg({f_2\over f}\bigg)^{(n-2)/\sigma}{{\Gamma{((n-\tau +1)/\sigma)}}\over {\Gamma{((3-\tau)/\sigma)}}}, \qquad n\ge 2,\label{fnf23}
\end{equation}
that is, if the conjectures (\ref {sigma},\,\ref {tau}) are used
\begin{eqnarray}
{f_n\over {f_2}}&=&(-1)^n\, {S_n\, }{\Gamma {(3n/2-2)}}, \label{fnf21}\\
S_n&=&\bigg({f_2\over f}\bigg)^{(3n/2-3)}.
\end{eqnarray}
This is the basic equation of our paper. It determines 
all the $f_n$'s, $n>2$ as a function of $f_2$ at the price of 
one free parameter $f$ only. 
We fitted Eq.(\ref {fnf21}) to the data \cite {jknum}, fixing $f_2$ at
the central value measured for each $q$=10, 15 and 20, $n$ varying from 3
to 8. We obtained a $\chi ^2/{d.o.f.}$ equal to 1.3 for 17 degrees of freedom  
with
\begin{equation} \label {f}
f=0.295\pm 0.003 \quad .
\end{equation}
The excellent agreement is 
illustrated in Fig.[1] where the numerical data \cite {jknum} for
$(-1)^n f_n/{(f_2\, S_n)}, \, n\le 8$ are compared to $\Gamma(3n/2-2)$ 
\,(continuous line) and plotted against the order $n$. Also indicated are 
the orders of magnitude estimated for $n=9$ and $10$.
\begin{figure}[ht]
\psfig{figure=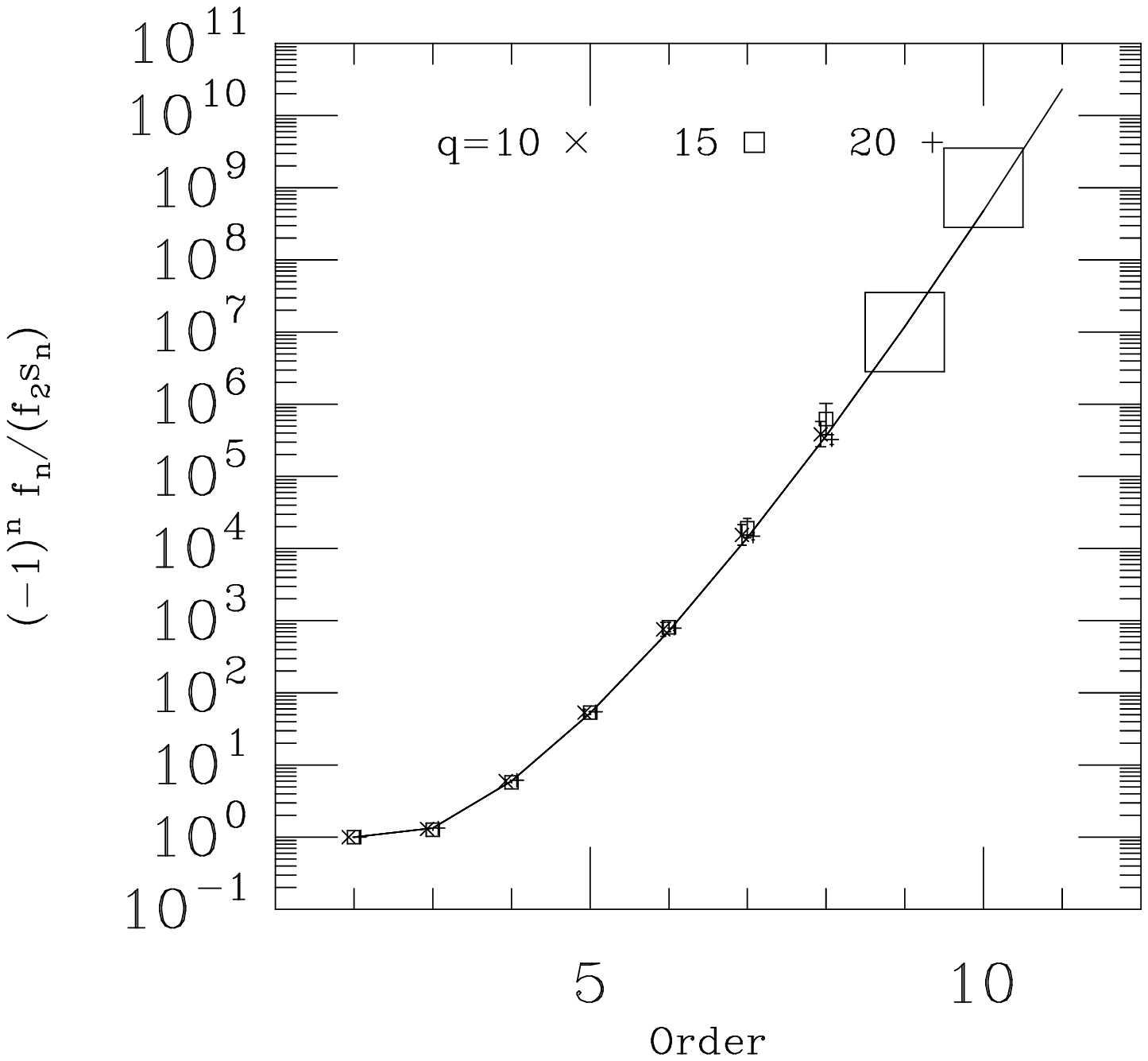}
\caption[1]{Comparison of the prediction of Eq.(\ref {fnf21}) 
(continuous line) with the numerical data of Ref.\cite {jknum}, versus
the order $n$. The large squared symbols at $n=9$ and $10$ indicate 
orders of magnitude only. The scale factor $S_n$, which contains 
the only free parameter $f$, exhausts the full $q$-dependence of 
the $f_n$'s.} 
\end{figure}
In order to appreciate whether the values (\ref {sigma}, \ref {tau})
of $\sigma$ and $\tau$ are actually requested by the numerical
data, we have repeated the fit letting also these parameters free.
The new fit requires 
\begin{eqnarray}
\sigma&=&0.67\pm 0.01, \label{sigmafit}\\
\tau&=&2.34\pm 0.10, \label{taufit}\\
f&=&0.28\pm 0.02,
\end{eqnarray}
with $\chi ^2/{d.o.f.}=1$ for 15 $d.o.f.$., in perfect
agreement with our conjecture. 

This nicely confirms the existence of a large domain of "low" $q$ values 
where the fluctuations of the system are strongly influenced by the
critical properties at $q_c$. The Fisher droplet model, supplemented
by our conjecture that the whole $q$-dependence is embedded in $f_2$,
gives them an economical and accurate description.  We note that the ansatz 
\cite {bhat3} predicts much smaller values for the highest order cumulants, 
missing $f_8$ by a factor about thirty. These results not only 
support the existence of an essential singularity in the free energy, 
but also show that this singularity {\it does lead to 
detectable effects}. We now proceed to construct the free energy explicitly .

\subsection{The Disordered Phase Free Energy}
The free energy can be reconstructed either directly from the droplet 
formulation (\ref {droplet}), replacing the sum by an integral as done in 
\cite{langer}, or from the asymptotic series (\ref {series}) with $f_n$ 
given by (\ref {fnf23}) or (\ref{fnf21}) and the
$\Gamma$-function replaced by its integral representation (Borel
resummation). In view of the results of the previous subsection, it is natural
to introduce the following rescaled free
energy $\phi$, expressed as a function of a rescaled temperature $z$, 
\begin{eqnarray}
\phi (z)&\equiv&\left (\frac{f_2}{f}\right )^{2/\sigma}{1\over {f_2}}\,\Big (F(\beta)-F(\beta _t)-f_1(\beta -\beta _t)\Big ),\label{phi}
\\
z&\equiv&{-(\beta -\beta _t)}({f_2/f})^{1/\sigma}. \label{z}
\end{eqnarray}
With this choice of normalizations, we have $\phi (0)=\phi '(0)=0$ and
$\phi ''(0)=1$. For generic values of $\sigma$ and $\tau$, $\phi$ reads
\begin{equation}
\phi (z)={1\over {\Gamma((3-\tau)/\sigma)}}\int_0^\infty {dt\over {t^{(\sigma +\tau -1)/\sigma}}}e^{-t}\Big (e^{-zt^{1/\sigma}}-1+zt^{1/\sigma}\Big ), \label{phigen}
\end{equation}
and specializing to the values (\ref {sigma}, \ref {tau})
of $\sigma$ and $\tau$, we finally get
\begin{equation}
\phi (z)=\int_0^\infty {dt\over {t^3}}e^{-t}\Big (e^{-zt^{3/2}}-1+zt^{3/2}\Big ). \label{phimod}
\end{equation}
This function is obviously holomorphic for $Re\, z>0$, 
which includes the stability region $\beta <\beta _t$ of the disordered phase.
It can be analytically continued in the complex $z$-plane by
deforming the integration contour in t. The only singularity is a branch
point at $z=0$, easy to characterize. The
contribution to $\phi$ from any finite $t$ interval $[0,t_0]$ is an entire
function, as well as that from $t_0$ to $\infty$ of the linear part (in $z$) 
of the integrand. Hence the singular part of $\phi$ is that of
\begin{equation}
\phi (z)=\int_{t_0}^\infty {dt\over {t^3}}e^{-t-zt^{3/2}},
\label{phi32}
\end{equation}
for any positive $t_0$. Next, as $z$ is continued to $z=\vert z\vert exp(\pm i\pi)$, one may continuously move the contour $[t_0, +\infty]$ to $[t_0, \mp i\infty]$
leading to a discontinuity along the negative real axis
\begin{eqnarray}
\Delta&\equiv&\phi (-\vert z\vert  +i\epsilon)-\phi (-\vert z\vert -i\epsilon), \\
&=&-\int_{t_0-i\infty}^{t_0+i\infty}{dt\over {t^3}}e^{-t-zt^{3/2}}.\label{delta}
\end{eqnarray}
In the following, we define $\phi (z)$ by (\ref {phimod}) in
the whole complex plane cut along the negative real axis, on which
$\phi$ acquires an imaginary part.
For $\vert z\vert$ small, the above integral can be estimated by
steepest descent, giving
\begin{equation} \label{imphi}
Im\, \phi (-\vert z\vert +i\epsilon)= -\sqrt \pi \,{\Big ({3\vert z\vert \over
2}\Big )
^5} \,exp\Big (-{4\over{27z^2}}\Big ) \,I(z),
\end{equation}
where the function $I(z)$, easy to evaluate numerically, goes smoothly
to one as $z$ goes to $0$ along the negative real axis.
All this is quite similar to the analysis performed by Langer
\cite {langer} for the Ising case, where the magnetic field
plays the role of $z$. Whenever a precise evaluation of $\phi (z)$ is
needed, one can either numerically integrate the analytic continuation of 
(\ref {phimod}) using suitable contours, or use its large $\vert z\vert$
expansion, which is found to be:
\begin{eqnarray}
\phi (z)&=&\gamma /6-3/4-\log (z)/3-2z\sqrt \pi +S
\label{phiasym} \\
S&=&{2\over 3}z^{4/3}\sum _{q\geq 0,\ne 2}^\infty {(-1)^q\over {q!}}z^{-2q/3}\Gamma ({{2q-4}\over 3}),
\end{eqnarray}
where $\gamma$ is the Euler's constant. 
The derivation of this expansion can be obtained by a Sommerfeld-Watson
method. One starts from the formal series expansion of $\phi$ in $z^n$, 
truncated at some large $n$ value $N$. The sum $\sum _0^N(-1)^na_n$ is replaced by the integral $1/(2i)\oint d\nu\,a_\nu /\sin (\pi\,\nu)$ over a contour 
encircling all integers from $0$ to $N$. After sending $N$ to $\infty$ the
contour is deformed so as to  encircle the poles of the integrand 
situated on the negative real axis, and the residue formula is finally
applied. A similar expansion for the second derivative of $\phi$ with
respect to $z$ is much easier to derive starting from
\begin{eqnarray}
\phi ''(z)&=&\int _0^\infty dt\,{e^{-t-zt^{3/2}}},\\
&\equiv& z^{-2/3}\int _0^\infty dt\,{\exp({-t/z^{2/3}-t^{3/2})}},
\end{eqnarray}
and expanding the latter integral in powers of $z^{-2/3}$ 
which yields a convergent series in this variable \cite{navelet}.

As we have seen, the closed form (\ref {phimod}) of the {\it thermodynamical}
free energy incorporates the information acquired upon the $\sim 10$  
first energy cumulants in simulations performed on {\it large but finite 
lattices} during finite (Monte-Carlo) times. 
That it makes sense is justified by exact results \cite {borgskot}, which, in 
short, state that on a finite lattice, $1/A\,\log (Z^{(d)}_A)$ gives
the thermodynamical limit $F^{(d)}$ up to corrections which are 
exponentially  small in the linear size of the lattice. In contrast, the next
section will illustrate that the energy distribution $P_A(E)$ on a  
finite lattice exhibits strong finite size effects, specific to the 
singularity structure of the thermodynamical free energy, and
corroborated by a numerical simulation at $q=9$. The procedure followed 
to study $P_A$ is similar to that used in \cite {bhat3}. 

\section{Finite Size Effects and Scaling Properties in the Internal Energy Distribution}
\subsection{The Probability Density for the Energy}
Given the rescaled free energy $\phi (z)$, the probability density for observing a lattice averaged energy
density E in a supposedly disordered bulk phase at $\beta =\beta _t$,
on a lattice of area $A$, is obtained by inverse Laplace transform using
Eqs. (\ref{ZA}, \ref{pb}, \ref{FA}). We find
\begin{equation}  \label {pe}
P_{{\beta _t},A}(\epsilon)={\frac {A_r}{2i\pi }}{\int _{\bar{z}-i\infty }^{
\bar{z}+i\infty }}dz\exp {\left (A_r\, (\phi (z)-\epsilon \,z)\right )}, \\
\end{equation}
where the rescaled energy, area and inverse temperature $\epsilon , A_r$
and $z$ respectively are defined as
\begin{eqnarray}
\epsilon &=&{\frac {1}{f}}{\left (\frac {f_2}{f}\right )}^{1/2}(E-E^{(d)}),
\label {eps} \\
A_r&=&f{\left (\frac {f}{f_2}\right )^2}A, \label {ar} \\
z&=&-{\left (\frac {f_2}{f}\right )}^{3/2}(\beta -\beta _t).
\label{zz}
\end{eqnarray}
The symbol $E^{(d)}$ (or $E^{(o)}$) represents the exactly known internal
energy of the disordered (or ordered) phase in the thermodynamical limit 
\cite{wu}.
In the rest of this paper, and depending on the context, we will 
use either the physical $E,\,A$ and $\beta$ variables, or
their rescaled forms $\epsilon ,\,A_r$ and $z$. For economy of 
notations however, we will often keep the same name for functions of them.
The variable E (or $\epsilon$) is considered as continuous, 
$\beta$ (or $z$) as unlimited in the imaginary direction. 
Eq.(\ref {eps}) maps the physical region $E\in [-2,
0]$ over a large interval in $\epsilon$, the more so $f_2$ is large (see
Eq. (\ref{eps}), that is $q$ close to 4. 
Since on large lattices $P$ is very sharply peaked around $\epsilon 
=0$, it will be justified (and convenient) to consider also $\epsilon$ 
as unlimited, unless specified. Due to the analyticity properties of 
$\phi(z)$, the integral (\ref{pe}) does not depend on $\bar{z} \geq 0$. 

With the above prescriptions, one can verify that $P_{\beta _t,A}(\epsilon)$ 
is a probability density, normalized to one with respect to
integration over $\epsilon$, that $<\epsilon>=0$, $<\epsilon ^2>=1$,
and more generally that the cumulants of $\epsilon$ for the density
$P_{\beta _t,A}$ have exactly the values assigned by Eqs.(\ref{phi}, 
\ref {z}, \ref {fnf21}), a very powerful check for numerical integration 
of (\ref {pe}). Any explicit $q$-dependence has disappeared, and
the effective area $A_r$ is the only external parameter. Its relation (\ref{ar})
to the physical area $A$ predicts the scaling property that different
$q$-values lead to the same $P_{\beta _t,A}(\epsilon)$ if they are measured
on lattices of linear size proportional to $f_2$ (i.e. approximately 
to $\xi$ according to (\ref {f2}), which sounds reasonable).  
We note that the largest L values used in \cite {jknum}
for $q=[10,\,15,\,20]$ were precisely chosen roughly proportional
to $\xi$.

This scaling behaviour is independent of the particular form of
$\phi (z)$, and we turn to the study of those properties of the distribution 
which are specific to our construction. The results of this section are
all summarized through their application to the case $q=9$ with $L=$80,
120 and 200, shown in Fig.[2]. 
The rescaling of $A=L^2$ to $A_r$ (\ref{ar}) requires the knowledge
of $f_2$ at $q=9$. There the correlation length is 14.9
\cite {borgs}, and we evaluate $f_2$ from its value at $q=10$ \cite {jknum} 
by (\ref {f2}), finding $f_2(q=9)=12.8$.  
The quantity shown in Fig.[2] is
\begin{equation} \label {q}
q_{\beta _t,A} (\epsilon)={1\over {A_r}} \log \left (\frac {P_{\beta _t,A}(\epsilon)} {\sqrt {A_r}}\right )\quad . \\
\end{equation}
The continuous lines correspond to the numerical integration of (\ref{pe}), 
and the other ones to various analytical approximations described below. 

\subsection{The Distribution above the Disordered Peak, $\epsilon >0$. }
For $\epsilon \geq 0$, the saddle point equation for the integral 
(\ref {pe}), namely 
\begin{equation}
\frac {d\,\phi}{d\,z}=\epsilon, \label{zs}
\end{equation}
has a unique, non negative solution $z_s(\epsilon)$. Taking $\bar {z}=z_s$ in
the definition (\ref{pe}) makes the numerical integration easy. The saddle 
point estimate of $P_{\beta _t,A}$, valid at large $A_r$, is
\begin{equation} \label{saddle}
P_{\beta _t,\,A}^{(s)}=\sqrt  {\frac {A_r\,\phi ^{'\!'}(z)}{2\,\pi}} \exp {\Big (A_r\left [\phi (z)-z{\frac {d\,\phi}{d\,z}}\right ]\Big )},\quad
z=z_s(\epsilon),\quad \epsilon>0.
\end{equation}
In this approximation, the thermodynamical limit of $q_{\beta _t,A}$ (\ref{q}) 
exists for any non negative $\epsilon$, and defines, up to a linear
term in $\epsilon$, a non trivial, 
concave, microcanonical entropy density $S(E)$. Moreover,  
for $\epsilon$ large, an estimate $z_{as}$ of $z_s$ can be obtained, 
by solving Eq.(\ref{zs}) analytically when  
the two leading terms only of the large $\vert z\vert$ expansion 
(\ref {phiasym}) are kept. Then $\phi(z_{as})$ and $\phi ^{'\!'}(z_{as})$ are
computed at the same order, and plugged into (\ref{saddle}), avoiding any 
integration. The corresponding algebraic expressions 
are quite lengthy, and not  here. They are used for
Fig.[2], showing that the approximation is very good in practice
shortly after $\epsilon$=1. Farther away, these expressions show that   
$q_{\beta _t,\,\infty}(\epsilon)$ become  
proportional to $-(\epsilon+2\,\sqrt \pi)^4$. This means an extremely fast 
decrease of $P(\epsilon)$ on the right hand side of the peak, much faster 
than any Gaussian.

\subsection{The Distribution below the Disordered Peak. Scaling at $\epsilon 
<0$ }
The side $\epsilon <0$ is more unusual, its properties reflecting
the tendancy of the disordered system to order. As $\epsilon$ passes
through $0$, the solution of (\ref {zs}) disappears from the first
Rieman sheet, and we need to optimize the integration contour in
(\ref{pe}) differently. Given $\epsilon$, we move it to a contour 
$z\equiv x\pm i\,y(x,\epsilon)$ passing through $0$ and determined 
by the condition
\begin{equation} \label{im0}
Im\,(\phi -\epsilon \,z)=0.
\end{equation}
In practice, because $Im(\phi(x))$ is never large before the
integrand in(\ref{pe}) gets exponentially damped at large $\vert x\vert$, we
expand $\phi(x+i\,y)$ in $y$ to second order (note that $\phi $ is
analytic and thus expandable at any non zero x), and solve (\ref{im0}) for 
$y$ as a function of $\phi (x)$ and of its two first derivatives along the real
axis. Along this approximate path, the integrand does not 
oscillate too much,  allowing  smooth numerical integration, whose
result  is drawn as continuous lines in Fig.[2].

\begin{figure}[ht]
\psfig{figure=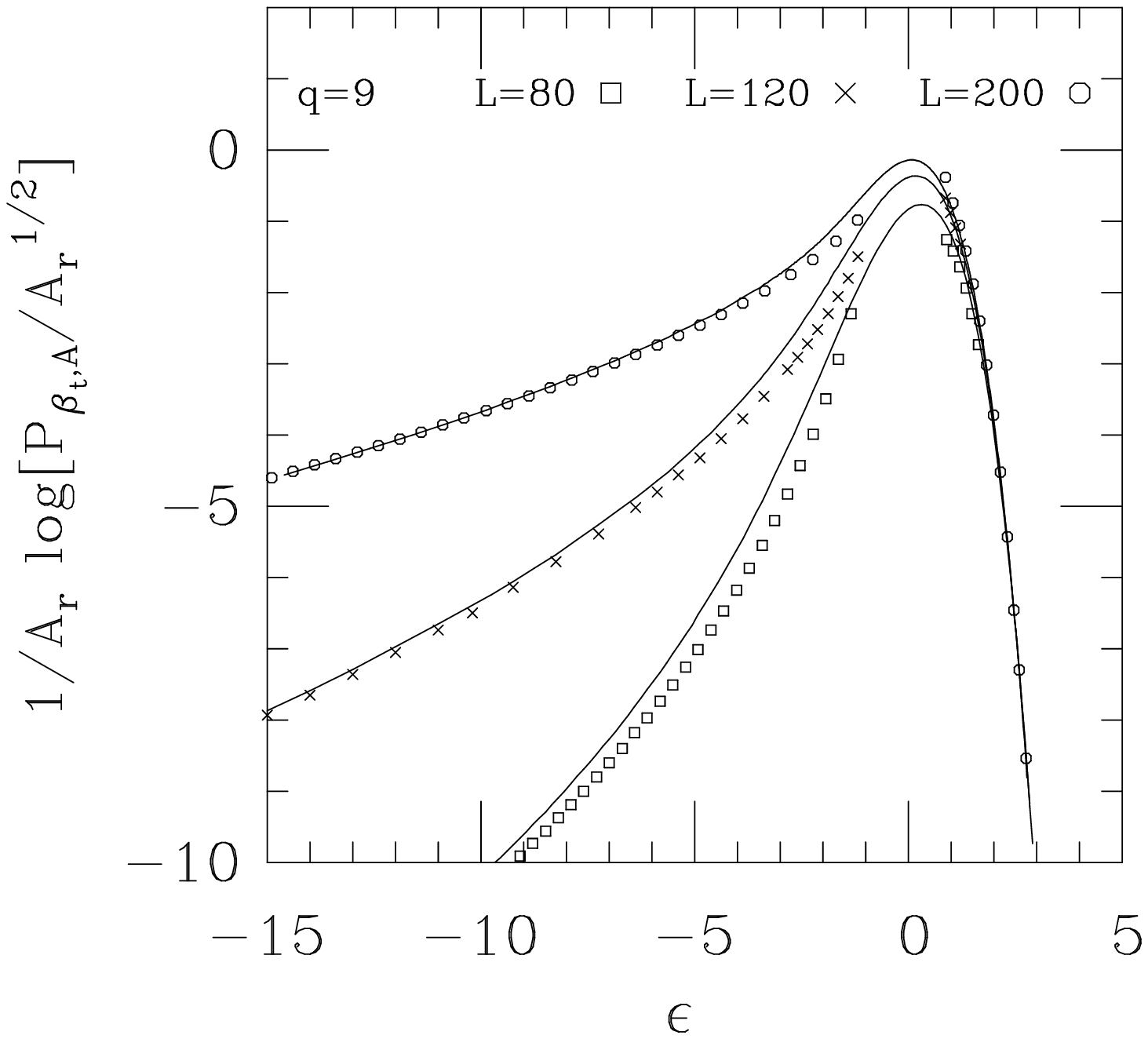}
\caption[2]{ The energy distribution of the disordered phase at $q=9$ as a 
function of $\epsilon$ for $L=80,\,120,\,200$. The continuous 
curves result from numerical integration of (\ref{pe}) and the symbols
represent analytical approximations (see text). As $L$ increases, the strong
broadening for $\epsilon <\, 0$ is a manifestation of the free
energy singularity at $\beta _t$.}
\end{figure}

We extended the curves down to unrealistically small probabilities, in
order to illustrate the peculiarities of the distribution, both in shape 
and in size dependence. Compared to the right hand side of the peak, the 
curve on the left is much broader and it {\it broadens} substantially as 
$L$ increases. A similar behaviour of the Ising magnetization distribution 
at zero field has been analyzed in \cite{schulman}.
We now correlate these features to the singular part of 
$\phi(z)$, via an analytical estimate of $P_{\beta _t,A}(\epsilon)$ 
valid at $\epsilon <0$. \\

The integral (\ref {pe}) is also equal to that of minus twice the imaginary 
part of the same integrand above the cut:
\begin{equation} \label{pen}
P_{{\beta _t},A}(\epsilon)={-\frac {A_r}{\pi }}{\int _{-\infty }^{0}}
dx\exp {\left (A_r\, (Re\,\phi (x)-\epsilon \,x)\right )}\,\sin(A_r\,Im\,\phi (x)). \\
\end{equation}
 For $-\epsilon$ not too small, the integrand is exponentially damped
when $\vert x\vert$ increases, the more so $A_r$ is large, while
it vanishes at small $\vert x\vert$ due to the behaviour (\ref {imphi})
of $Im\,\phi$. Thus for $A_r$ large, but fixed, and $\vert x\vert$ small
enough, $Im\,\phi$ can be replaced by its  approximation (\ref {imphi}) and  
$\sin (A_r Im\,\phi)$ by its argument.  The existence of a saddle point 
$x_s$ follows. Neglecting $Re\,\phi(x_s)$ in front of $\epsilon \, x_s$, one 
finds
\begin{eqnarray}
x_s&=&-\left (\frac {8}{27\,A_r\,\vert \epsilon \vert }\right )^{1/3},\label
{colneg}\\
\frac {1}{A_r^2}\,P_{\beta _t,A}&=&{\frac {2}{3}}\,{(A_r\vert \epsilon \vert)}^{-7/3}\,\exp\left (-(A_r\,\vert \epsilon \vert  )^{2/3}\right )\,
,\quad \epsilon <0. \label {peneg}
\end{eqnarray}

Hence we have the remarkable scaling prediction that, up to a factor
$A^2$, the distribution depends on both the energy and the size through
the single variable $A_r\,\vert \epsilon \vert$. Moreover, the
 argument of the exponential damping of the probability {\it is not
extensive} and its $\epsilon$-dependence {\it is less than linear}. These
two properties explain the observed broadening of the distribution, and
their very derivation emphasizes the role of a non vanishing imaginary part.
Fig.[2] shows that the approximation (\ref{peneg}) works quantitatively,
the more so the product $\vert A_r\epsilon \vert$ is large, which
actually insures that $x_s$ and thus $A_r Im\,\phi$ are small. The
behaviour (\ref{peneg}) will be important for our discussion of
metastability (see Section (5)).

\subsection{A Numerical Simulation at q=9}
We have just seen that the scaling law (\ref{peneg}) sets in for 
$A_r\,\vert \epsilon \vert $ sufficiently large. Hence a numerical 
simulation designed to check it should accumulate reasonable
statistics for events whose probability relative to the peak value is 
quite small. This means very extensive simulations, typically
of the size of those of Refs.\cite {jknum, billoire}. However, 
Fig.[2] shows that substantial finite size
effects already appear in $q_{\beta _t,A}(\epsilon)$
, Eq.(\ref {q}), in a crossover region in between $\epsilon
=0$ and the above mentioned scaling domain. In order
to complement the comparison with simulations made for the cumulants (Section 
(3.1)), by a direct study of finite size effects
in energy distributions, we performed a medium size 
simulation for $q=9$, where the correlation length is about 1.4 larger
than at $q=10$, and on lattices of sizes $L=80, 120,$ and 200.

We used a Glauber type of algorithm. The site to  
be updated is chosen at random and its new spin value  
determined according to its Boltzmann weight.
At $\beta =\beta_t = \log(4)$ we performed runs consisting for each $L$ of
about $10^6$ sets of $L^2$ random hits at the lattice sites.  
We measured the internal energy density $E$ every 10 sets, thus
collecting a sample of $\simeq 10^5$ configuration energies for each L value.
Starting from a disordered configuration (spin chosen at random on each site) 
, the histogram obtained is proportional to the normalized probability
density $P_{\beta _t,A}$ discussed above. For comparing the
numerical data to the model distribution, Fig.(3), we divide each of them   
by its value at a reference energy which we choose to be
$E_0=-1$, and as a function of the internal energy $E$ we
plot the quantity 
\begin{eqnarray} \label {pe050}
p_{\beta _t,A}(E)&=&10^3\,\frac{1}{A}\,\log\left (\frac{P_{\beta _t,A}(E)}{P_{\beta
_t,A}(E_0)}\right ). \\
\end{eqnarray}
Here we restored the physical variables using Eqs.(\ref{eps},\,\ref{ar})
with $f_2=12.8$ as explained in Section (4.3).

At each value of $E$, the error quoted is estimated 
from the fluctuations around the mean observed 
in subsamples of $10^3$ to $10^4$ measurements. In this range, the error was 
found to depend weakly upon the bin size, in agreement with an estimated
autocorrelation length of order $10^2$ for the average energy (i.e. in
the peak region; it increases away of it). For the clarity of the figure, 
we suppressed the data points for which the statistics is too small to 
bring any useful information.

\begin{figure}[ht]
\psfig{figure=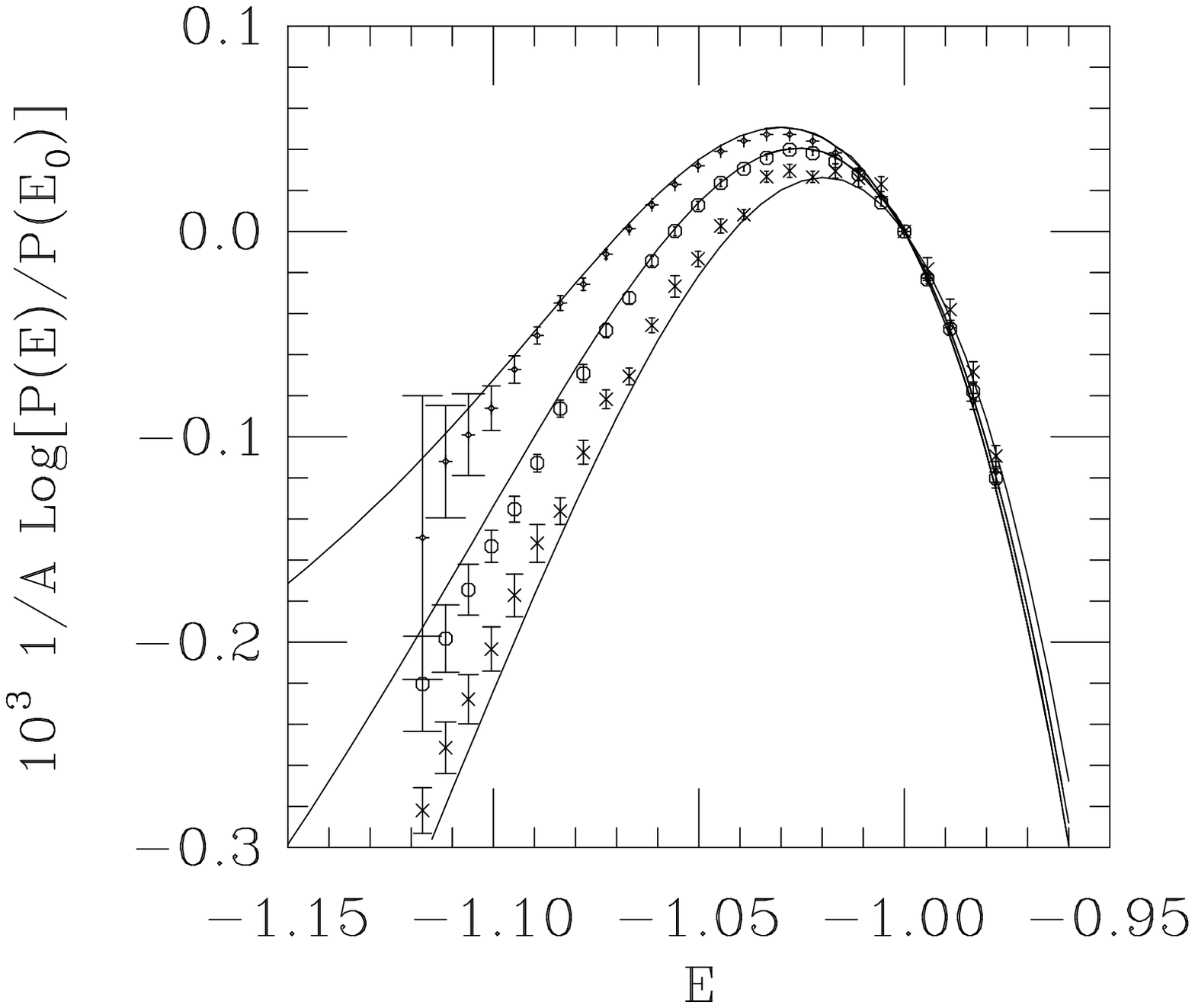}
\caption[3]{Comparison, for $p_{\beta _t,A}(E)$, Eq.(\ref {pe050}),
between the model predictions and numerical data taken at $q=9$, 
\,$L$=80, 120, and 200 (from bottom to top).}
\end{figure}

We see on Fig.[3] that both the shape in energy and the dependence in
size observed in the simulation are correctly reproduced by our construction, 
an absolute prediction since the only parameter $f$ 
was previously fixed from the study in section (3.1) of the cumulants 
at $q\ge 10$. There may be a tendancy for the size dependence to be too
strong in the model. We do not consider it as significant enough to
justify a retuning of $f$ or $f_2$, the more so we did not include in the
errors shown that due to the fluctuations at the reference energy
$E_0$. This error induces an overall uncertainty in the relative
vertical positions of the curves. We cannot exclude either that finite size
effects, in the sense of a residual $L$ dependence {\it in the free energy}, 
are still present at the lowest $L$.

We conclude that the adequacy of the model to predict quantitatively
the strong size dependence observed in the disordered phase energy 
distribution below the peak, as well as its (nearly) size independent
shape above the peak, constitutes a good evidence in favour of the analytical
structure of the free energy inherent to the droplet picture.

\section{Beyond the transition temperature. Metastability}

Metastability of the disordered phase refers to the possibility for the
system to stay disordered above $\beta _t$. 
Standard phenomenology \`a la Van der Waals, or Landau mean field theory, 
would state that the thermodynamical average energy density, which below
$\beta _t$ coincides with $<E>^d\,(\beta)=-d/d\beta \,F^{(d)}(\beta)$,
can be continued up to a so-called spinodal value $\beta _{sp}$,
separating a metastable from an unstable region. 

In the droplet picture, as well as in field theoretic
approaches, the point $\beta _t$ is a branch point,
and analytic continuation above $\beta _t$ is ill-defined.
An important question, not considered in this paper,
concerns the associated dynamics.  An extensive review 
can be found in \cite {gunton}; see also \cite{dynamics}.  This field is 
still subject to active research in the context of Ising-like or 
liquid/vapour transitions. After Langer \cite{langer, langer2} and followers, 
the nucleation rate of a metastable state is proportional to the imaginary 
part of the free energy along the cut. Most of the recent results concern
the Ising case (See for example \cite{penrose, schonmann, shlosman}). To our knowledge, these
dynamical aspects have not been studied for the Potts model above
$q_c$. Here we limit ourselves to a discussion of which thermodynamical
properties can be assigned the metastable states, starting from these
features of energy distributions which we found in Section (4) as
specifically due to the singular structure of the free energy. For the
 Ising case, related considerations can be found in \cite{schulman}.

\subsection{Energy Distribution away from $\beta _t$ by Reweighting}
On a finite lattice, 
the extrapolation of a given energy distribution from $\beta _t$ to
{\it any} (larger or smaller) $\beta$ {\it is unambiguously 
defined} by the corresponding reweighting of the Boltzmann factor. Hence, 
{\it if $P_{\beta _t,A}\,(E)$ is the pure disordered phase distribution}, 
then the (unnormalized) distribution at $\beta$ is (we
write it both in physical and rescaled variables (\ref {eps},\,\ref {ar},\,\ref {zz})
\begin{eqnarray} \label {dbe}
D_{\beta ,A}(E)&=&\exp (-A\,(E-E^{(d)})\,(\beta -\beta _t))\,P_{\beta _t,A}(E).\\
D_{\beta ,A}(\epsilon)&=&\exp (A_r\,\epsilon\,z)\,P_{\beta _t,A}(\epsilon).
\end{eqnarray}
\begin{figure}[ht]
\psfig{figure=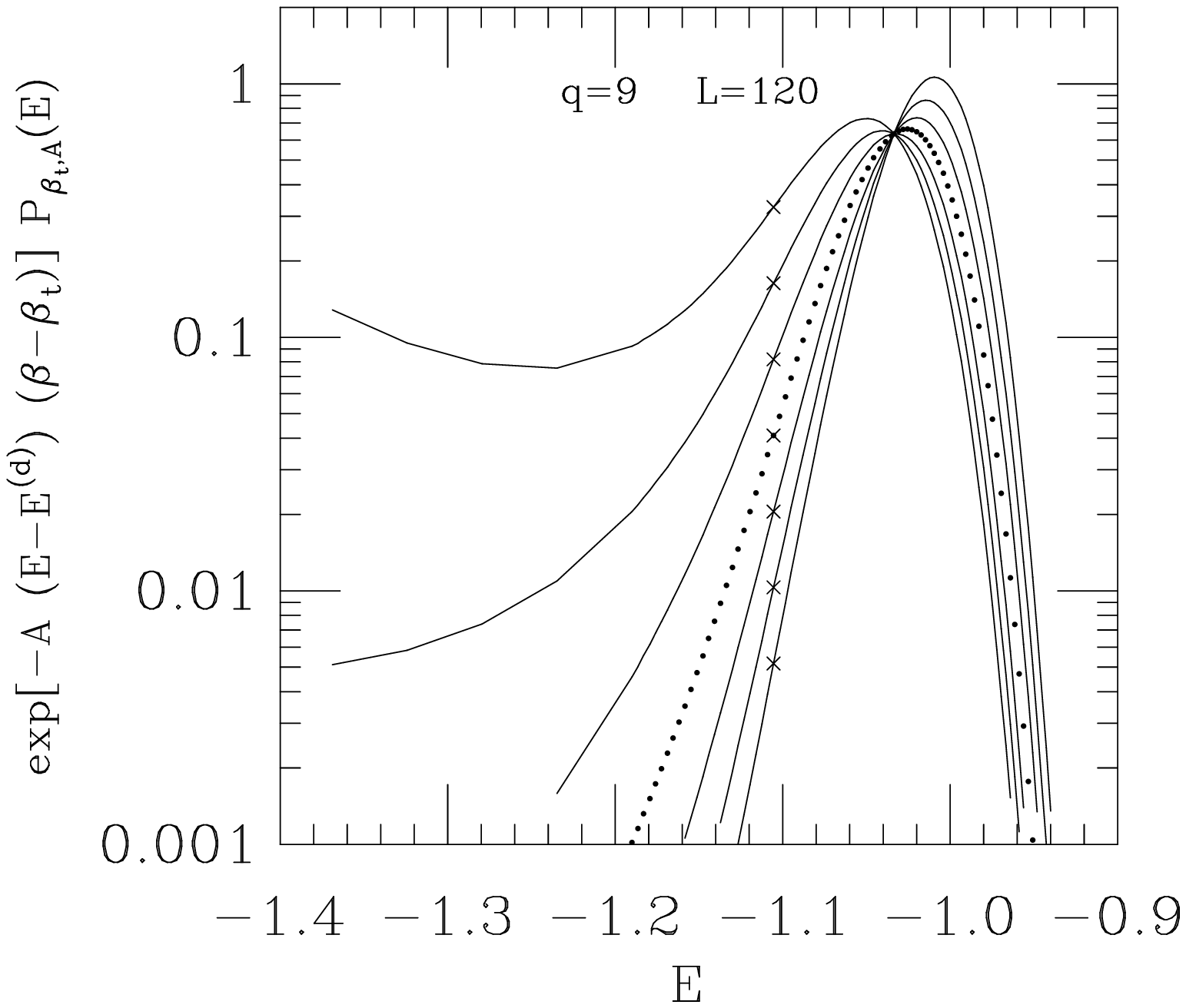}
\caption[4]{For the disordered phase at $q=9,\, L=120$, energy distributions for $(\beta -\beta _t)$ varying from $-2 \,10^{-3}$ to $2 \,10^{-3}$.  
On the left of the figure, temperature decreases from the bottom
to the top curves (overcooling), the dotted curve corresponding to $\beta =\beta _t$. The crosses correspond to the
inflexion points which all occur at the same $E^*=-1.104$.
As the temperature is lowered below the transition, a minimum of the 
distribution enters the graph from the left.}
\end{figure}
The reweighting (\ref {dbe}) is often used to smoothly interpolate, or slightly extrapolate numerical data taken at some fixed 
$\beta $ point (here at $\beta _t$). The results of \cite {jknum} 
show that it is feasible to perform a high statistics simulation at 
$\beta _t$ without tunneling from one phase to another, so that an accurate
{\it numerical} determination of the pure phase $P_{\beta _t,A}\,(E)$ 
was obtained. 

We got this distribution for finite lattices from a {\it theoretical} 
ansatz for the thermodynamical limit of the free energy, and now get 
its continuation beyond $\beta _t$ by (\ref {dbe}). 
The generic shapes of the reweighted distributions as $\beta$ is varied 
are shown in Fig.[4] for $q=9$ and $L=120$. We 
comment them in the light of the results of Section (4).
For $\epsilon $ positive, $P_{\beta _t,A}$ decreases
faster than a Gaussian (section 4.2), and the only effect of reweighting there
is to displace the peak position, whatever the sign of $z$ is.
For $\epsilon$ and $z$ both negative (see sections 4.3 and 4.4), the 
exponential growth of the Boltzmann factor eventually   
wins over the smoother behaviour (\ref{peneg}) of $P_{\beta _t,A}$.
Hence, under overcooling the distribution finally
blows up for $\vert z\vert$ large enough, leading to a minimum at $E=E_m$,
visible in the upper curve on the left of Fig.[4], which 
exhibits the cubic like shape typical of a metastable situation.
In between $E_m$ and the location $E_M$ of the maximum,
all curves have an inflexion point at the
same value $E=E^*$: It is the inflexion point of the (finite size)
microcanonical entropy $S_A(E)$.

The depth of the minimum relative to the peak height is a measure of
the barrier to be crossed by the system in order to flip to its stable
state, and thus controls the lifetime of the metastable state.
There is no meaning in considering the distribution (\ref{dbe}) below 
$E_m$, where for the least the contributions of the ordered states, 
obtainable from ours by duality, and of mixed phase configurations 
\cite {billoire} must be included. As a {\it definition}, we take the
distribution (\ref{dbe}) cut below $E_m$ as representative of the 
metastable disordered phase. 

On a given lattice, the barrier disappears when $\beta$ becomes
equal to $\beta ^*$, the slope  at $E^*$ of
\begin{equation}
d_{\beta ,A}\,(E)\equiv\frac {1}{A}\,Log(D_{\beta ,A}\,(E)). \\
\end{equation}
The size dependent point $(\beta ^*,\,E^*)_A$, which plays the role 
of the classical spinodal point, is thus determined by:
\begin{eqnarray} 
\frac {\partial ^2}{\partial E^2}\,d_{\beta ,A}\,(E)_{\vert
_{E=E^*}}&=&0, \label {espinode} \\
\frac {\partial}{\partial E}\,d_{\beta ,A}\,(E)_{\vert
_{E=E^*}} &=&\beta ^* -\beta \quad . \label {bspinode}
\end{eqnarray}
At $L$=120 and $q$=9, again chosen for illustration, we find from the model 
distribution at $\beta _t$ (the dotted curve of Fig.[4]) 
\begin{eqnarray} 
E^*&=&-1.104  \label{estar}\\
\beta ^*&=&\beta _t+3.44\, 10^{-3}=1.38973 \quad.  \label{betastar} 
\end{eqnarray}
Our statistics at $\beta _t$ (Section (4.4)) are not
sufficient to give significant results by reweighting, and  we explored
the above issue by new simulations in the vicinity of $\beta _t$. 
For each $\beta$ above $\beta _t$, where flips from the
$d$ to the $o$ phase may become frequent, we performed ten independent runs,
each run being ten times shorter than at $\beta _t$. Flips, seen as
jumps of the lattice average energy density $\bar {E}_L$, do occur, but
the time histories show that the configurations with $\bar 
{E}_L$  larger than $\sim$ -1.24 can be safely labelled as disordered.
Although an accurate determination of $E^*$
from the data is not possible , we consistently find that the distribution 
$d_{\beta ,A}\,(E)$ is essentially a straight line around  $E=-1.1$, and
we measure its slope (\ref {bspinode}) there.
Our numerical results as a function of
$\beta$ are reported in Fig.[5], and compared with the
prediction (\ref {bspinode}) with $\beta ^*$ fixed by the
model (\ref {betastar}). The agreement is significant: the spinodal
point location could be identified numerically as the value of $\beta$ where
data points for the left hand side of (\ref {bspinode})
linearly extrapolate to 0.
\begin{figure}[ht]
\psfig{figure=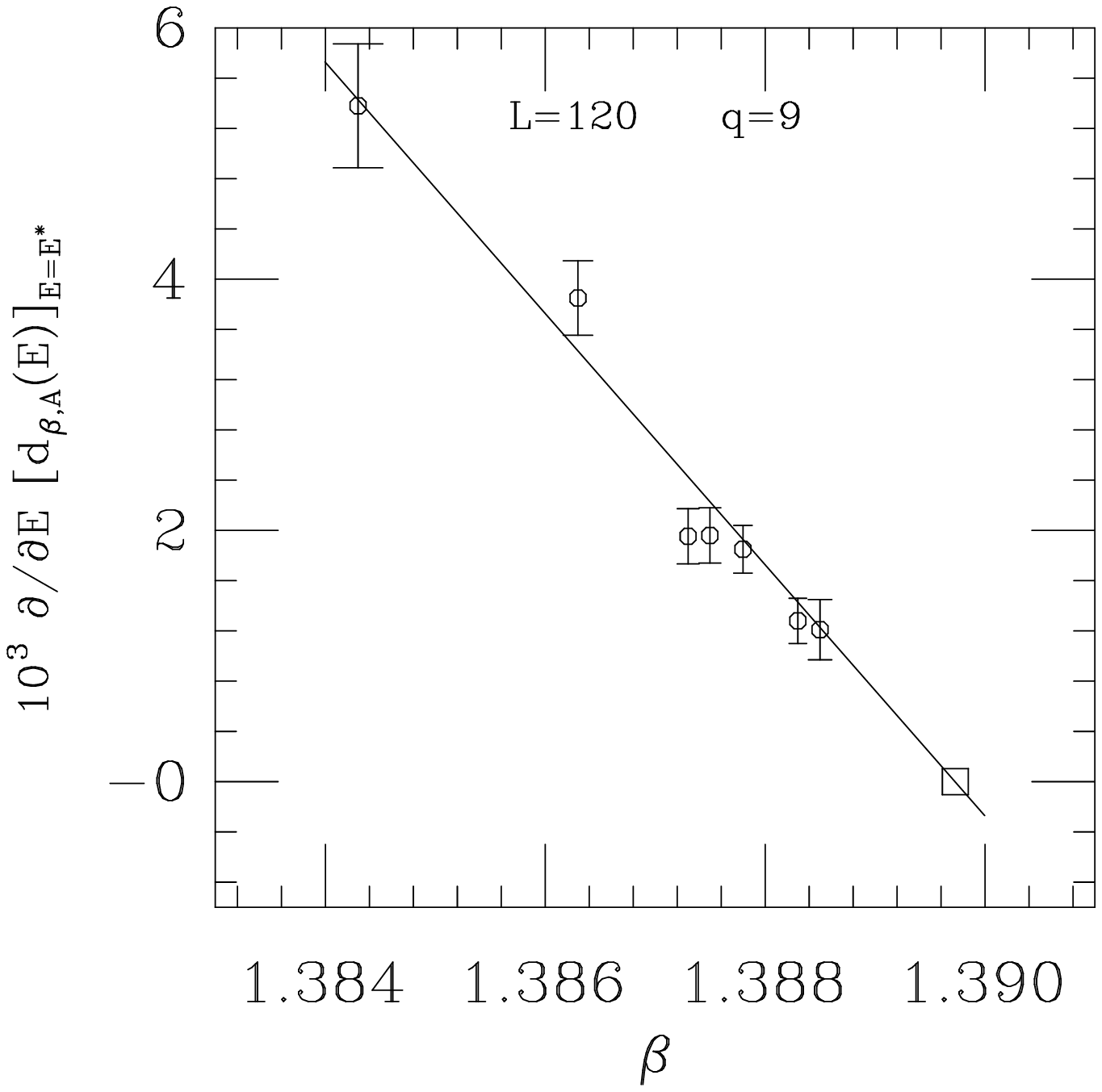}
\caption[5]{The slope $\frac {\partial}{\partial E}\,d_{\beta
,A}\,(E)_{\vert _{E=E^*}}$ as a function of $\beta$ from
numerical simulations ($\circ$). The straight line represents
Eq. (\ref {bspinode}), with $\beta ^*$ given by the model 
(\ref{betastar}): the slope vanishes at the spinodal value 
$\beta ^*$ ($\Box$).}
\end{figure}

We do not repeat the whole analysis for other lattice sizes. The
important point is that the slope of $d_{\beta ,A}\,(E)$ in the relevant 
energy region {\it decreases as $L$ increases}, as seen on
Fig.[3]. Our discussion of Section (4.3) interpreted this
size dependent behaviour of the energy distribution  in terms of the
singularity structure of the free energy. In particular, 
Eq. (\ref {peneg}) indicates that, as $A\to \infty$,  
$d_{\beta ,A}\,(E)$ as well as its derivative eventually goes to 0, at least
for any $E$ held fixed sufficiently far below the peak. Via Eq. 
(\ref {bspinode}), this in turn implies that $\beta ^*(A)\to \beta _t$ when
$A\to \infty$. Hence we conclude that the metastability
interval $\beta ^*-\beta _t$ shrinks to zero in the thermodynamical limit,
reproducing the result of the Maxwell construction applied to
Van der Waals theory.
\subsection{Free Energy for a Finite System in a Metastable State}

We have defined the disordered energy distribution up to $\beta ^*$ 
as the distribution (\ref {dbe}) cut below its minimum at
$E_m$. It is then natural to use it in the standard relations (\ref {ZA},
\ref{pb}, \ref{FA}) to associate a free energy $F_A(\beta )$ to the
metastable disordered phase. So for $\beta \leq \beta ^*$, we set
\begin{eqnarray} 
\frac {Z_A(\beta )}{Z_A(\beta _t)}&\equiv& 
\exp(A\,(F _A(\beta )-F_A(\beta _t))=\sum _{E}\,^{'}\, D_{\beta
,A}(E),\label{extraz} \\
\phi _A (z)&\equiv&\left (\frac{f_2}{f}\right )^{3}{1\over {f_2}}\,\Big (F_A(\beta)-F(\beta _t)-f_1(\beta -\beta _t)\Big ).\label{phia} 
\end{eqnarray}
In the above equations $\sum _{E}\,^{'}$ means summation above 
$max\, [E_m,\,-2]$. Note the index $A$ in $F_A$ and $\phi _A$, 
to be explained soon. For consistency of course, 
one must find that, for $z\geq 0$,
$\phi _A$ coincides with $\phi$, which by construction  
is $A$ independent. This can be checked by plugging in the
representation (\ref{pe}) in the reweighted distribution (\ref{dbe}), 
and the result in (\ref{extraz}, \ref{phia}), leading to
\begin{equation} \label {phiaa}
\exp(A_r\,\phi _A(z))={\frac {A_r}{2\, \pi}}\int _{\epsilon _m}^{\infty}d\epsilon \,\int _{-\infty}^{+\infty}d\rho \,\exp \left [A_r(\phi (\bar{z} +i\rho)\,-\epsilon \,(\bar {z}-z+i\rho))\right ].
\end{equation}
The lower bound $\epsilon _m$ corresponds to the cut $E _m$, 
when necessary. For $z\geq 0$, one may choose $\bar {z}=z$, $\epsilon _m$ 
can be set to 
$-\infty$ with an exponentially small error and integration over
$\epsilon$ yields the desired result. 

The choice $\bar {z}=z$ cannot be done for $z^*<z<0$, where $z^*$ corresponds to
$\beta ^*$ via (\ref{zz}), due to the branch point at $z=0$. 
Furthermore, the cut at $\epsilon _m$ becomes
relevant. For both reasons, $\phi _A$ as defined by (\ref
{phia}) actually becomes A-dependent. We illustrate this
definition of the free energy for metastable states by the
continuous line of Fig.[6], drawn for $L=120$ and $q=9$. Also
shown is the result of the integral (\ref{phia}) with $\epsilon _m$ 
replaced by a {\it fixed} cut off (here corresponding to a cut at 
$E_{cut}$=-1.24). For completness, we also plot the real part of the
input $\phi (z)$ (\ref {phimod}) for $z<0$ 
(dashed line). All definitions provide the same result for $z\geq 0$.

\begin{figure}[ht]
\psfig{figure=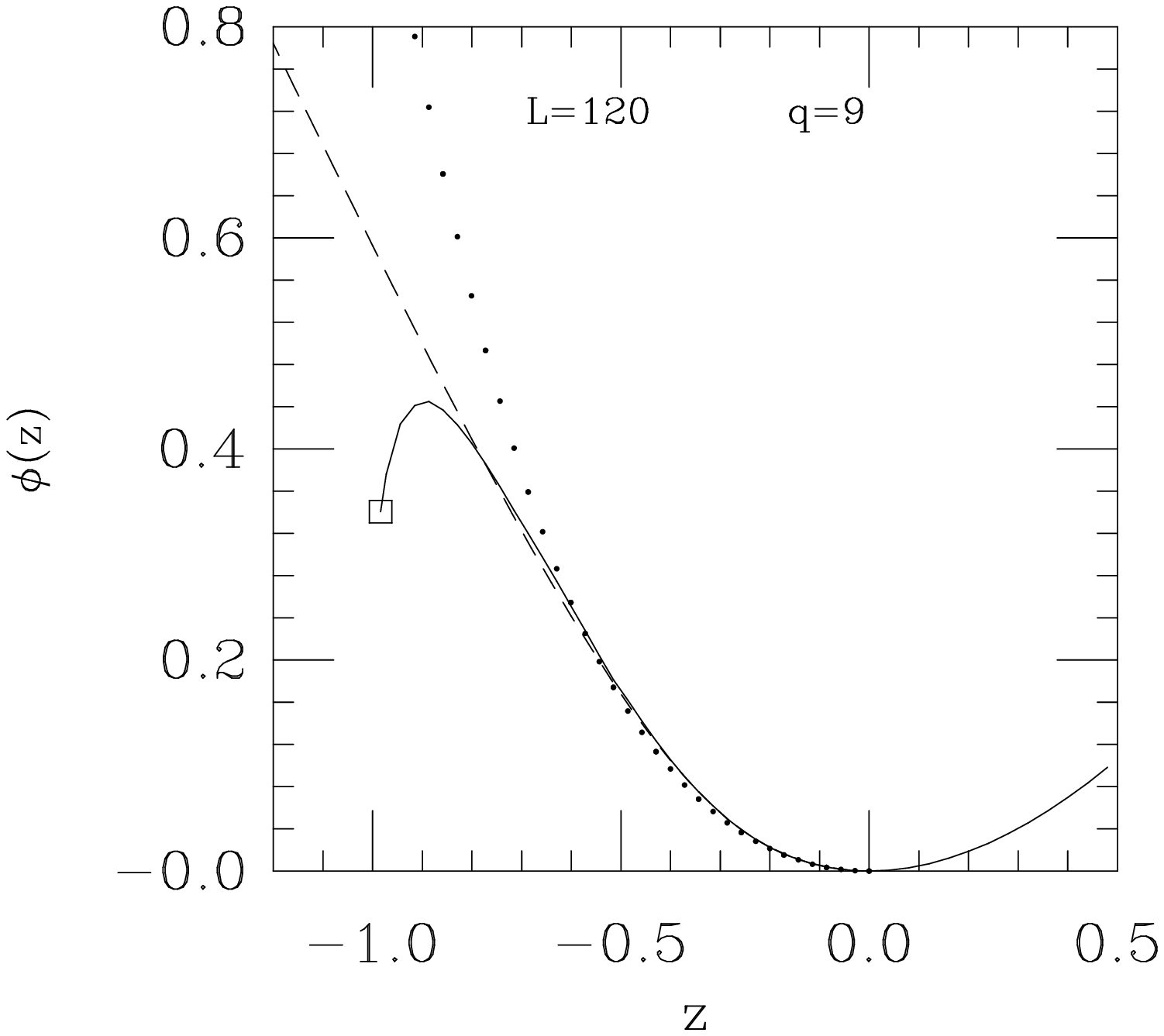}
\caption[6]{The rescaled free energy for metastable 
states ($z\leq 0$). Are shown the sum (\ref{phia}) where the cut 
$\epsilon _m$ is the $z$ dependent minimum of the reweighted 
distribution (continuous curve),
the same sum  with a fixed cut ($E_{cut}$=-1.24, dots), and 
the real part of the input $\phi (z)$ (dashed curve). 
The symbol $\Box$ marks the end point of the metastability region.
It moves right and down with increasing $L$, ending at the origin
in the thermodynamical limit.
All definitions give the same answer $\phi (z)$ for $z\geq 0$.}
\end{figure}

Let us add a few comments. First one observes that $\phi _A$ and
Real\,$\phi$ are quite similar just below $z=0$. We know that Im\,$\phi$
is very small there, or equivalently, that the droplet expansion (\ref
{droplet}) can be safely troncated at the size $\ell _c$ where 
the term of order $\ell$ is minimum \cite {langer,isakov,penrose}.
Near the end point $z^*$, one then notice the intriguing fact that $\phi _A$ 
is not convex. However, because the restriction on the sum 
(\ref {extraz}) depends on $\beta$ (via $E_m$), the energy cumulants do 
not coincide with the successive derivatives of $F _A(\beta )$, they must 
be computed by actual averages over the distribution (\ref {dbe}).
In particular, the quadratic fluctuations of the energy, of course always 
positive, are not proportional to $\phi _A^{''}$, which thus may be
negative. 

Likewise, the fluctuation-dissipation theorem does not hold
and the specific heat, $C=-\beta ^2\,d/d\beta \,<E>$ is  proportional
neither to the quadratic fluctuations of the energy, nor to the second
derivative of $\phi _A$. Averaging the energy, we compute $<E>$ 
and find it has a minimum near $\beta ^*$, showing that 
$C$ (after a maximum) does become negative. Such a phenomenon disappears in
the thermodynamical limit since, in this limit, $\beta ^*\to \beta _t$. 
If a fixed cut is set (dots in Fig. (5)), one recovers the familiar 
connection between energy cumulants and derivatives of the free energy, 
and $\phi$ is convex. Our choice of cutting at $E_m$ looks physically 
sensible. On a finite lattice, it leads to a situation similar to that
of Van der Waals theory. In the thermodynamical limit, it defines a
{\it disordered microcanonical entropy density} $S^{(d)}(E)$, defined
and concave for $E^{(d)}\leq E\leq0$. A similar construction of its
ordered counterpart $S^{(o)}(E)$ is immediate through duality for
$-2\leq E\leq E^{(o)}$. The entropy $S(E)$ for the whole system is
finally achieved as usual by supplementing $S^{(o)}(E)$ and $S^{(d)}(E)$ by
their common tangent straight line between $E^{(o)}$ and $E^{(d)}$. In
this way, our approach provides a model entropy $S(E)$ for any $E$.

\section{Summary and Conclusions}

We have applied a Fisher's version of the droplet picture to the $2D$
$q$-states Potts model in the vicinity of the temperature
driven transition, which is first order above $q_c=4$. 
The droplet parameters were constrained by requiring that critical 
properties of the free energy are recovered as $q\to q_c$, complying
with Fisher's scaling ideas and previous empirical observations based on 
large $q$ expansions of the energy cumulants \cite {bhat1,bhat2,bhat3}.
Using duality of the Potts model, we focussed on the disordered
phase properties and showed that, at the price of a unique $q$-independent 
free scale, the energy cumulants of order $\leq$10 measured in a
numerical simulation by Janke and Kappler were accurately reproduced (Fig.[1]).
The analytical structure of the thermodynamical free energy in the inverse 
temperature $\beta$ plane was then analyzed, and a branch point essential 
singularity at the transition fully characterized. Next we investigated the 
consequences of these properties directly on the energy distribution  
currently generated in numerical simulations on finite lattices. In 
particular, evaluating this distribution analytically far
away from its peak, we established a direct link between its size
dependence there and the free energy discontinuity around the branch
cut. Finite size effects are also present close to the peak of the
distribution and in good agreement with the result of a numerical
simulation which we performed at $q=9$ for comparison. We finally 
discussed static aspects of metastability. We defined a free energy 
beyond the transition point $\beta _t$ via the reweighting of the energy 
distribution at finite size, showing that, again due to the essential 
singularity, this can be done up to a size dependent spinodal point only, 
which moreover coincides with $\beta _t$ in the thermodynamical limit.

We noted earlier that various versions of the generic droplet picture
have been used to study the first order transition of the Ising model
below its critical temperature, and the associated issues of
singularity of the free energy, metastability, and nucleation rates.
As already mentioned in Section (2.3), the Fisher's version which we
apply as been questionned for the Ising case in that it assigns a
droplet an interfacial energy which grows faster than the
perimeter ($\sigma>1/2$ in (\ref{droplet})), which sounds odd at least 
for large droplets. Binder has shown \cite{binder, binderrev} how scaling 
prescriptions valid close to the critical point ($T=T_c,\, h=0$ for Ising) 
can reconcile the Fisher's description and the original
geometric picture of nucleation. We expect that similar arguments 
apply to the Potts case around $q=q_c,\, \beta =\beta _t$. In any case, 
our work shows that following the Fisher's point of view provides an 
extremely efficient and economical parametrization of the pure phase free 
energies, as particularly illustrated by its adequacy to describe the
ten first energy cumulants: a)~their measurements in numerical
simulations \cite{jknum} {\it require the predicted, non-geometrical,
values $\sigma =2/3$ and $\tau =7/3$} (see Eqs.(\ref{sigmafit},
\ref{taufit})), b) the pseudo interfacial tension $\omega$ scales
as prescribed by the Widom exponent for the conventional tension.

We see three directions in which it would be interesting to pursue with
the Potts model, noting that the above indices differ from
their geometrical values more than they do for the Ising case. 
i) Derivation of exact results, in analogy with what has
been done for Ising or Ising-like models on cluster distributions, 
large order cumulants, metastable states. ii) Dynamics at and
near the transition
and relationship between nucleation rates and singularities of
the free energies. iii) Relationship between finite size properties of 
energy distributions at the transition and the discontinuity of the
pure phase free energy. The existence of such a connection is demonstrated
by our derivation of Eq.(\ref{peneg}) (which can be easily
extended to arbitrary values of the indices $\sigma$ and $\tau$). 
Accordingly, high statistics numerical data \cite{jknum, 
billoire} may provide a direct access to the free energy 
discontinuities, without requiring heavy dynamical
investigations.

\vspace{0.5cm}

\centerline{\bf Acknowledgments}

\vspace{0.5cm}

It is a pleasure to thank R.Balian, A.Billoire, T.Garel, W.Janke and
R.Lacaze for their interest in this work and stimulating discussions.


\begin{thebibliography} {40}
\bibitem{isakov} S.N. Isakov, Commun. Math. Phys. {\bf 95} (1984) 427.
\bibitem{fisher}
M.E.Fisher, {\it Physics} \rm (N.Y.) {\bf 3} (1967) 255; in {\it Critical
Phenomena, Proc. of the Int. School of Physics "Enrico Fermi", Course
LI}, ed. M.S.Green, (Academic Press,1971) 1.
\bibitem{mayer} J.E. Mayer and M.G. Mayer, {\it Statistical Mechanics},
Chapter {\bf 14}, John Wiley, New York (1940).
\bibitem{langer}
J.S.Langer, {\it Annals of Physics} {\bf 41} (1967) 108.
\bibitem{binder} K.Binder, {\it Ann. Phys. }{\bf 98} (1976) 390.
\bibitem{delyon} F.Delyon,{\it J. Stat. Phys. }{\bf 21} (1979) 727.
\bibitem{aizenmann} M.Aizenmann, F.Delyon and B.Souillard,
{\it J. Stat. Phys. }{\bf 23} (1980) 267.
\bibitem{gunton} J.D.Gunton, M.San Miguel and P.S.Sahni, The Dynamics
of First-order Phase Transitions, in {\it Phase Transitions and
Critical Phenomena}, Eds. C.Domb and J.L.Lebowitz, Vol. {\bf 8} (Academic
Press, 1983).
\bibitem {binderrev} K.Binder, {\it Rep. Prog. Phys.} {\bf 50} (1987)
783.
\bibitem{schulman} L.S.Schulman, in Ed. V.Privman {\it Finite Size
Scaling and Numerical Simulation of Statistical Systems}, World
Scientific, Singapore, 1990, p. 489, and references therein.
\bibitem{dynamics} {\it Dynamics of First Order Transitions}, HLRZ, KFA
J\"ulich Workshop, Ed. H.J.Herrmann, W.Janke and F.Karsch (World
Scientific, Singapore, 1992).
\bibitem{abraham}
D.B.Abraham and P.J.Upton, {\it Int. Journal of Mod. Phys.} {\bf C3}
(1992) 1071.
\bibitem{rikvold}P.A.Rikvold and B.M.Gorman, Ed. D.Stauffer, {\it Annual
Review of Computational Physics}, Vol. {\bf I}, World Scientific, Singapore,
1994.
\bibitem{gunther}C.C.A.Guenther, P.A.Rikvold and M.A.Novotny, {\it
Physica} {\bf A 212} (1994) 194.
\bibitem{acharyya} Recent studies of nucleation include: V.Cataudella, G.Franzese, M.Nicodemi, A.Scala and
C.Coniglio, {\it Phys. Rev.} {\bf E 54} (1996) 175; M.Acharyya and D.Stauffer, {\it Eur. Phys. J.} {\bf B5} (1998) 571; C.S.Schioppa, F.Sciortino and
P.Tartaglia, {\it Phys. Rev.} {\bf E 57} (1998) 3797; L.Bocquet,
F.Restagno and T.Biben, cond-mat/9901180. For a field
theoretic approach: J.Alexandre, V.Branchina and J.Polonyi, {\it Phys.
Letters} {\bf B 445} (1998) 351; A.Strumia, N.Tetradis and C.Wetterich,
hep-ph/9808263.
\bibitem{potts} R.~B.~Potts, Proc. Camb. Phil. Soc. {\bf 48} (1952) 106.
\bibitem{jknum}
W.Janke and S.Kappler, {\sl J. Phys.\/} (France) {\bf I7}(1997) 663.
\bibitem{billoire} A. Billoire, T. Neuhaus and B.A. Berg {\it Nucl.
Phys. } {413} (1994) 795. This paper contains references to earlier
attempts at interface tension determinations.
\bibitem{bhat1} T. Bhattacharya, R. Lacaze and A. Morel,
{\sl Europhys. Lett.\/} {\bf 23} (1993) 547; {\sl  Nucl. Phys. B\/} (Proc. Suppl.) {\bf 34} (1994) 671.
\bibitem{bhat2} T. Bhattacharya, R. Lacaze and A. Morel,
 {\sl J.Phys.\/}  (France) {\bf I7} (1997) 81.
\bibitem{bhat3} T. Bhattacharya, R. Lacaze and A. Morel,
{\sl  Nucl. Phys.} {\bf B435\/} (1995) 526.
\bibitem{cardy} J.Cardy, Renormalisation Group Theory of Branching
Potts Interfaces, cond-mat/9806098.
\bibitem{wu}
F.Y. Wu, {\sl Rev. Mod. Phys.\/} {\bf 54} (1982)
235.
\bibitem{baxter}
R.J. Baxter, {\sl J. Phys.\/} {\bf C6} (1973)
L-445; {\sl J. Stat. Phys.\/} {\bf 9} (1973) 145.
\bibitem{hiley} B.J. Hiley and M.F.Sykes, {\it J. Chem. Phys.} {\bf 34}
(1961) 1531.
\bibitem{essam} J.W.Essam and M.E. Fisher, {\it J. Chem. Phys.} {\bf 38} (1963) 802; M.E.Fisher and M.F. Sykes, {\it Phys. Rev.} {\bf 114} (1974) 45;
N.J.Guenther, D.A.Nicole and D.J.Wallace, {\it J. Phys. A} {\bf 13}
 (1980) 1755.
\bibitem{kasteleyn} P.~W.~Kasteleyn and C.~M.~Fortuin, {\it J. Phys. Soc. Japan}
 {\bf 26} (Suppl.), 11 (1969).
\bibitem{arisue} H.Arisue and K.Kabata, The large $q$ expansion of the
energy cumulants for the two-dimensional $q$-state Potts model,
hep-lat/9810029.
\bibitem{xi} A. Kl\"umper, A. Schadschneider and J.
Zittarz, {\sl Z. Phys.\/} {\bf B76} (1989) 247; A. Kl\"umper,
 {\sl Int. Journal of Mod. Phys.\/} {\bf B4} (1990) 871;
E. Buffenoir and S. Wallon, {\sl J.Phys.\/} {\bf A26} (1993) 3045.
\bibitem{borgs} C. Borgs and W. Janke, {\sl J. Phys.\/}
(France) {\bf I2} (1992) 2011; W.Janke and S.Kappler, {\sl Nucl.
Phys.\/}(Proc. Suppl.) {\bf B42} (1995) 770; {Europhys. Lett.} {\bf 31}
(1995) 345.
\bibitem{widom} B. Widom, {\it J. Chem. Phys.} {\bf 43} (1965) 3892.
\bibitem{navelet}The function $\phi ^{"}$ can also be calculated as a 
Meijer's function (H.Navelet, private communication). Similar functions are
studied in \cite{kastrup}.
\bibitem{kastrup}H.A.Kastrup, cond-mat/9803269. 
\bibitem{borgskot} C.Borgs and R.Koteck\'y, {\sl J. Stat. Phys.\/} 
{\bf 61} (1990) 79; C.Borgs, R.Koteck\'y and S.Miracle-Sole,
{\sl J. Stat. Phys.\/} {\bf 62} (1991) 529.
\bibitem{langer2} J.S.Langer {\it Ann. Phys.} {\bf 54}(1969) 258.
\bibitem{penrose} O.Penrose, {\it J. Stat. Phys.}{\bf 78} (1995) 267.
\bibitem{schonmann} R.H.Schonmann and S.B.Shlosman, {\it Comm. Math.
Phys.} {\bf 194} (1998) 389.
\bibitem{shlosman} S.B.Shlosman, in {\it Statphys20}, Eds. A.Gervois et al.,
 Elsevier (North Holland) 1999.
\end{thebibliography}
\end{document}